\pdfoutput=1

\documentclass[twocolumn,english]{revtex4}
\usepackage{times,amssymb,amsmath,graphicx}
\usepackage[bookmarks=false,pdffitwindow=false,pdfstartview={FitH}]{hyperref}
\usepackage[T1]{fontenc}
\usepackage{babel,multirow}
\usepackage{color}

\begin{document}

%%  ======================================================================  %%
%%  TO JEST  >> W A Z <<  O DLUGOSCI 70+8 ZNAKOW I MA SIE MIESCIC W LINII   %%
%%  ======================================================================  %%

\title{
  Sub-Sharvin conductance and Josephson effect in graphene 
}

\author{Adam Rycerz\footnote{Correspondence: 
  \href{mailto:rycerz@th.if.uj.edu.pl}{rycerz@th.if.uj.edu.pl}.}}
\affiliation{Institute for Theoretical Physics,
  Jagiellonian University, \L{}ojasiewicza 11, PL--30348 Krak\'{o}w, Poland}

\date{February 26, 2026}

\begin{abstract} 
Titov and Beenakker 
[\href{https://doi.org/10.1103/PhysRevB.74.041401}{%
Phys.\ Rev.\ B {\bf 74}, 041401(R) (2006)}]  
found, by solving the Dirac-Bogoliubov-De-Gennes equation,
that the product of critical current and normal-state resistance for
superconductor-graphene-superconductor (S-g-S) Josephson junction 
takes values (for a short junction and zero temperature) between
$I_cR_N\approx{}2.1$ and $I_cR_N\approx{}2.4$
in units of $e/\Delta_0$, where $\Delta_0$ is the superconducting gap. 
These values are notably higher than the tunnelling
bound ($\pi/2$), but lower than the ballistic bound ($\pi$). 
Here we analyze numerically the tunneling of Cooper pairs through S-g-S
junctions in which the longitudinal electrostatic potential profile is tuned,
within gates electrodes, from a~rectangular to a~parabolic one.
In the unipolar regime (i.e., when the chemical potential is above the top of
a~barrier, $\mu>0$), it is found that $I_cR_N$ gradually evolves
from the graphene-specific to the ballistic value.
At the same time, the normal-state conductance increases from the sub-Sharvin
value of $1/R_N\approx(\pi/4)\,G_{\rm Sharvin}$ towards to the 
Sharvin value $G_{\rm Sharvin}=g_0|\mu|W/(\pi\hbar{}v_F)$, with
the conductance quantum $g_0=4e^2/h$, the junction width $W$, and the
Fermi velocity in graphene $v_F$. 
In contrast, in the tripolar regime ($\mu<0$), both normal-state conductance
and the critical current are suppressed when smoothing the potential; 
however, $I_c{}R_N$ remains close to the graphene-specific range,
even for a~parabolic potential.
The skewness of the current-phase relation is also discussed. 
\end{abstract}

\maketitle

%%%%%%%%%%%%%%%%%%%%%%%%%%%%%%%%%%%%%%%%%%%%%%%%%%%%%%%%%%%%%%%%%%%%%%%%%%%%%%
%%%%%%%%%%%%%%%%%%%%%%%%%%%%%%%%%%%%%%%%%%%%%%%%%%%%%%%%%%%%%%%%%%%%%%%%%%%%%%

\section{Introduction}

Soon after the theoretical prediction \cite{Jos62} and experimental
confirmation \cite{And63,Jos64,Lik79} of the Josephson effect,
a~current-bias Josepson tunnel junction was used to demonstrate that 
a~macroscopic variable, namely the phase difference across a~junction,
follows the rules of quantum mechanics \cite{Mar85,Dev85}.
Since early 2000s, superconducting circuits containing Josephson junctions
are often regarded as the most promissing condensed-matter platform for
multiqubit quantum information processing \cite{You05,Wen17}.
Both the tunneling- \cite{Jos64} and weak-link \cite{Lik79} Josephson
junctions offers possibility to control the system parameters by
a~bias current or magnetic field \cite{Tin04}.
However, a~control via the electrostatic field effect, commonly used in
semiconductor-based devices, is widely-considered asfar more versatile and
practical \cite{Aka94,Mor98,Hee07,Sim19,Gol21}.
In particular, two superconductors separated by a~two-dimensional system,
such as a~semiconducting heterostructure \cite{Aka94},
thin Au film \cite{Mor98}, or graphene \cite{Hee07}, allow one to modify
the critical current ($I_c$)  elecrostatically, using gate electrode. 
From a~physical perspecive, such Josephson-Field-Effect-Transitors operate
since---in the short-junction limit---the product $I_cR_N$ (where $R_N$
denotes the resistance of the junction in the normal state) saturates at
a~value of order $\Delta_0/e$ determined by the excitation gap $\Delta_0$ 
in the superconductors (with a~geometry-dependent prefactor)
\cite{Bee92,Tit06,Mog06}. 
Therefore, the electrostatic control over $R_N$ implies analogous control
over $I_c$.

For a~superconductor-graphene-superconductor (S-g-S) Josephson junction,
it was shown that the dimensionless prefactor relating $I_cR_N$ and
$\Delta_0/e$ varies within the range of $2.1-2.4$
\cite{Tit06,Mog06,Eng16,Nan17}.
This range is significantly narrower than that allowed for generic
mesoscopic Josephson junctions at zero temperature
\cite{Bee92,Amb63,Kul75}, i.e., 
\begin{equation} 
\frac{\pi}{2}\leqslant{}I_cR_N
\left(\frac{e}{\Delta_0}\right)\leqslant{\pi}, 
\end{equation}
where the lower (upper) bound corresponds to the tunneling (ballistic)
limit.
What is more, the current-phase relation for S-g-S Josephson junction is
forward skewed with respect to the familiar sinusoidal behavior
\cite{Eng16,Nan17}, and the quantity 
\begin{equation}
\label{skewdef}
S= \frac{2\theta_c}{\pi}-1, 
\end{equation}
where $\theta_c$ is the phase difference corresponding to the maximum current
($I_c$), varies from $S\approx{}0.25$ to $S\approx{}0.42$, comparing to
the range of $0\leqslant{}S\leqslant{}1$ for generic mesoscopic Josephson
junctions.

\begin{figure*}[t]
\includegraphics[width=0.6\linewidth]{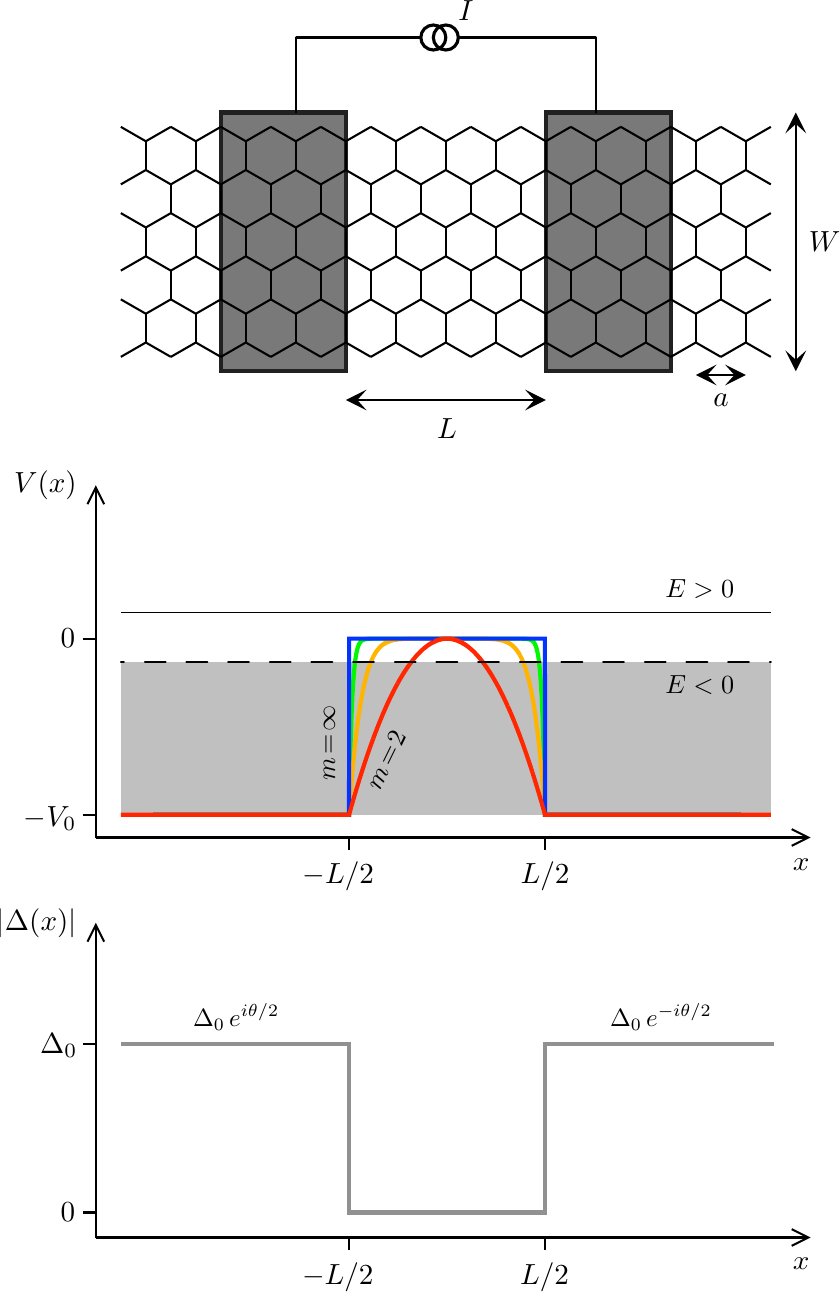}
\caption{ \label{setupjos}
  Top: Schematic of a~graphene strip of width $W$, contacted by two
  superconducting electrodes (dark areas) at a distance $L$.
  A~current source drives a~dissipationless supercurrent through
  the central region. A~separate gate electrode (not shown) allows us to
  tune the carrier concentration around the neutrality point.
  The lattice parameter $a=0.246\,$nm is also shown.
  Middle: Electrostatic potential profiles given by Eq.\ (\ref{vxmpot}) with
  $m=2$, $8$, $32$, and $m=\infty$ (i.e., the rectangular barrier).
  The Fermi energy $E$ is defined with respect to the top of a barrier.
  $E>0$ corresponds to unipolar n-n-n doping in the device;
  for $E<0$, a~tripolar n-p-n structure is formed.
  Bottom: Absolute value of the pair potential with the superconducting
  phases $\theta/2$ for $x<-L/2$ (left electrode) and $-\theta/2$ for $x>L/2$
  (right electrode). 
}
\end{figure*}

In this paper, we investigate numerically how the critical current and
skewness of the current-phase relation for S-g-S Josephson junction 
behave as functions of doping, supposing that the
electrostatic potential barrier is smooth (i.e., potential varies
slowly on the scale of atomic separation).
Similar problems
were addressed previously \cite{Tit06,Mog06}, but here we focus on
the effects of the potential profile, which is gradually tuned
from a parabolic to a rectangular shape (see Fig.\ \ref{setupjos}),
on the selected measurable quantities.
Thus, the paper complements recent studies on the normal metal-graphene-normal
metal (N-g-N) setup \cite{Ryc21b,Ryc21a,Ryc25}, in which analogous effects
of smooth barriers on conductance, shot noise power, and thermoelectric
properties were discussed. 

We notice that the effects of time-dependent fields, including Josephson 
quantum pumping, are beyond the scope of this work \cite{Acc25}.

The paper is organized as follows.
In Sec.\ \ref{modmet} we present the details of our numerical approach.
The key results for a~rectangular barrier are summarized in
Sec.\ \ref{recbar}.
Central results of the paper, concerning the normal-state conductance,
the critical current and skewness of the current-phase relation for smooth
potentials, are presented in Sec.\ \ref{smopot}.
The conclusions are given in Sec.\ \ref{conclu}.

\section{Model and methods}
\label{modmet}

We start from the Dirac-Bogoliubov-De-Gennes (DBdG) for the system
shown schematically in Fig.\ \ref{setupjos} \cite{Tit06,Mog06}, 
\begin{equation}
\begin{pmatrix}
 \mathcal{H}_0-\mu & \Delta \\
 \Delta^\star & \mu-\overline{\mathcal{H}}_0
\end{pmatrix}
\begin{pmatrix}
 \Psi_e \\
 \Psi_h
\end{pmatrix} 
=
\varepsilon\begin{pmatrix}
 \Psi_e \\
 \Psi_h
\end{pmatrix}.  
\end{equation}
Here
$\mathcal{H}_0=v_F\,\mbox{\boldmath$p$}\cdot\mbox{\boldmath$\sigma$} + V(x)$
is the Dirac Hamiltonian for $K$ valley, with
$v_F=\sqrt{3}\,t_0a/(2\hbar)\approx{}10^6\,$m$/$s the
energy-independent Fermi velocity in graphene ($t_0=2.7\,$eV is 
the nearest-neighbor hopping integral and $a=0.246$ is the lattice parameter), 
$\mbox{\boldmath$p$}=(p_x,p_y)$ the in-plane momentum operator 
(with $p_j=-i\hbar{}\partial_j$), $\mbox{\boldmath$\sigma$}=(\sigma_x,\sigma_y)$
(with $\sigma_j$ being the Pauli matrices),
$\Psi_e$ and $\Psi_h$ are the electron and hole wave functions, 
$\varepsilon>0$ is the excitation energy, and $\mu$ is the chemical potential
(hereinafter, $\mu=E$ --- the Fermi energy --- since the $T=0$ case is
considered) in the normal region measured with respect to the Dirac point,
so that $\mu=0$ corresponds to undoped graphene.
In the absence of a magnetic field, the Hamiltonian is time-reversal
invariant, 
$\overline{\mathcal{H}}_0
=\mathcal{T}\mathcal{H}_0\mathcal{T}^{-1}=\mathcal{H}_0$,
with $\mathcal{T}$ the time-reversal operator \cite{Bee06}. 

The complex pair potential $\Delta$ depends only the position along the
strip ($x$), and can be truncated by adopting the step-function model
for the two interfaces between normal region and superconductors 
at $x=-L/2$ and $x=L/2$, namely
\begin{equation}
  \Delta(x) = 
  \begin{cases}
    \,\Delta_0e^{i\theta/2}  &  \text{if }\ x < -L/2, \\
    \,0  &  \text{if }\ |x| \leqslant L/2, \\
    \,\Delta_0e^{-i\theta/2}  &  \text{if }\ x > L/2,   
  \end{cases}
\end{equation}
with the bulk superconducting gap $\Delta_0$ and the phase difference
between the superconductors $\theta$. 

The electrostatic potential energy is chosen as 
\begin{equation}
\label{vxmpot}
  V(x) = -V_0\times
  \begin{cases}
    \,1  &  \text{if } |x| > L/2, \\
    \,\left|2x/L\right|^m  &  \text{if } |x| \leqslant L/2, 
  \end{cases}
\end{equation}
such that changing the value of $m$ tunes the potential from
a parabolic shape ($m=2$) to rectangular shape ($m=\infty$).
The potential $V(x)$ is therefore continuous and constant in the leads
($x<-L/2$ or $x>L/2$).

It was shown in Refs.\ \cite{Tit06,Mog06} by analysing the spectrum of
Andreev states for $\varepsilon<\Delta_0$ that for the short-junction limit, 
i.e., $L\ll{}\xi_0$ (with the superconducting coherence length
$\xi_0=\hbar{}v_F/\Delta_0$; for instance, $\xi_0\approx{}550\,$nm
for superconducting electrodes made with molybdenum rhenium \cite{Nan17}),
the Josephson current can be written as
\begin{equation}
\label{ijoth}
  I(\theta)=\frac{e\Delta_0}{\hbar}\sum_{n=0}^{N-1}
  \frac{T_n\sin\theta}{\sqrt{1-T_n\sin^2(\theta/2)}},  
\end{equation}
while the normal-state resistance is given by
\begin{equation}
\label{rnlan}
  R_N^{-1}=\frac{4e^2}{h}\sum_{n=0}^{N-1}T_n.
\end{equation}
In effect, both quantities are determined by the transmission probabilities
$T_n$ characterising a~graphene strip between two electrodes in the normal
state ($\Delta_0=0$). 
Eqs.\ (\ref{ijoth}) and (\ref{rnlan}) coincide, respectively, 
with the multichannel mesoscopic Josephson equation \cite{Bee92} and
the Landauer-B\"{u}ttiker formula \cite{But85};
both formulas are multiplied by a factor of two due to the additional
(valley) degeneracy in graphene.

The transmission probabilities in Eqs.\ (\ref{ijoth}) and (\ref{rnlan}),
each of which is attributed to the $n$-th normal mode, with the transverse
momentum $k+y=k_y^{(n)}=\pi{}(n+\frac{1}{2})/W$ (we assume the infinite-mass
boundary conditions, see Ref.\ \cite{Ber87}), can be found by solving the
scattering problem for the Dirac equation $\mathcal{H}_0\Psi=E\Psi$.
Taking the wave function in a form $\Psi = \phi(x)e^{ik_yy}$, with
$\phi(x) = (\phi_a,\phi_b)^T$, one obtains for the central region ($|x|<L/2$)
\begin{align}
  \phi_a' &= k_y\phi_a+i\frac{E-V(x)}{\hbar{}v_F}\phi_b,
  \label{phapri} \\
  \phi_b' &= i\frac{E-V(x)}{\hbar{}v_F}\phi_a-k_y\phi_b.
  \label{phbpri} 
\end{align}

For the leads ($|x|>L/2$), we have a~constant $V(x)=-V_0$, and the solutions
can be found analytically, for $E>-V_0$,
\begin{equation}
  \label{leadspm}
  \phi^{(+)}=\
  \begin{pmatrix} 1 \\ e^{i\vartheta} \end{pmatrix}
  e^{iK_x{}x},
  \ \ \ \ 
  \phi^{(-)}=
  \begin{pmatrix} 1 \\ -e^{-i\vartheta} \end{pmatrix}
  e^{-iK_x{}x}, 
\end{equation}
where $e^{i\vartheta}=(K_x+ik_y)/K$, $K =(E+V_0)/\hbar{}v_F$, and
$K_x=\sqrt{K^2-k_y^2}$.
The solutions $\phi^{(+)}$ and $\phi^{(-)}$ represent the propagating waves
going to the left and to the right (respectively).
The number solutions (for one direction of propagation) is given by
\begin{equation}
\label{noofmods}
  N=\lfloor{}WK/\pi\rfloor. 
\end{equation}

The mode-matching conditions for the interfaces at $x=-L/2$ and $x=L/2$
brought us to the linear system of equations (for a~given $E$ and $k_y$) 
\cite{Ryc21b}
\begin{widetext}
\begin{equation}
  \label{lsysrt}
  \left[
    \begin{matrix}
      \phi_a^{(-)}(x_1) & -\phi_a^{(1)}(x_1)  & -\phi_a^{(2)}(x_1) & 0 \\
      \phi_b^{(-)}(x_1) & -\phi_b^{(1)}(x_1) & -\phi_b^{(2)}(x_1) & 0 \\
      0 &  -\phi_a^{(1)}(x_2)  & -\phi_a^{(2)}(x_2)  & \phi_a^{(+)}(x_2) \\
      0 &  -\phi_b^{(1)}(x_2)  & -\phi_b^{(2)}(x_2)  & \phi_b^{(+)}(x_2) \\
    \end{matrix}
  \right]
  \left[
    \begin{matrix}
      r \\ A \\ B \\ t \\
    \end{matrix}
  \right]
  =
  \left[
    \begin{matrix}
      -\phi_a^{(+)}(x_1) \\
      -\phi_b^{(+)}(x_1) \\
      0 \\
      0 \\
    \end{matrix}
  \right], 
\end{equation}
\end{widetext}
where we have denoted $x_{1,2}=\mp{}L/2$, $r$ ($\,t\,$) is the reflection
(transmission)
amplitude asumming the scattering from the left to the right, and $A$, $B$
are arbitrary complex coefficients associated with
linearly-independent solutions $\phi^{(1)}$, $\phi^{(2)}$ of
Eqs.\ (\ref{phapri}) and (\ref{phbpri}) for the central area ($x_1<x<x_2$). 
In practice, we take $\left.\phi^{(1,2)}\right|_{x=x_1}=(1,\pm{}1)^T$ and
find $\left.\phi^{(1,2)}\right|_{x=x_2}$ by integrating Eqs.\ (\ref{phapri})
and (\ref{phbpri}) within a~standard forth-order Runge-Kutta algorithm.
(The details of the calculations will be given later.) 

Solving Eq.\ (\ref{lsysrt}), one finds the transmission amplitude $t$
for a~given $k_y=k_y^{(n)}$ and $E=\mu$, 
and the corresponding transmission probability $T_n(\mu)=T_{k_y}(E)=|t|^2$.
The supercurrent $I(\theta)$ and the normal-state resistance $R_N$ can
then be determined from Eqs.\ (\ref{ijoth}) and (\ref{rnlan}) 
by summing over $N$ modes, see Eq.\ (\ref{noofmods}).

\begin{figure*}[t]
\includegraphics[width=0.6\linewidth]{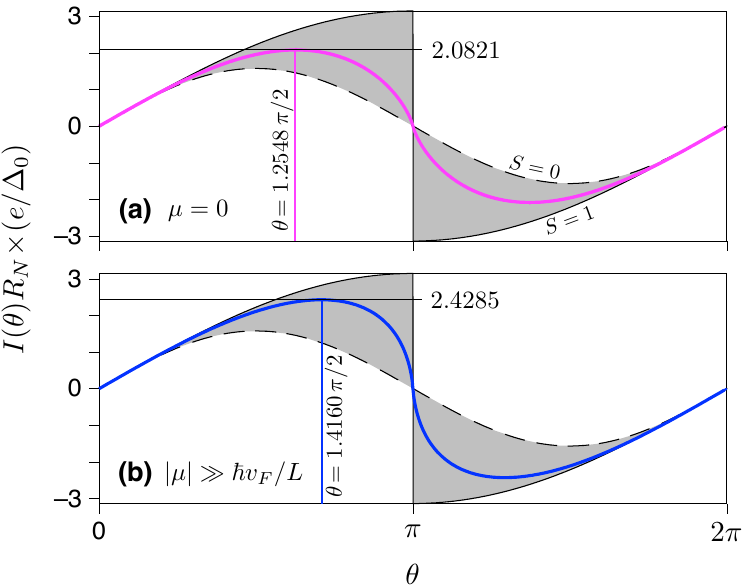}
\caption{ \label{ithrect1ab}
  Current-phase relation for S-g-S Josephson junction in the case of
  rectangular potential barrier and infinitely-doped leads,
  corresponding to 
  $m\rightarrow\infty$ and $V_0\rightarrow\infty$ in Eq.\ (\ref{vxmpot}).
  (a) The Dirac point ($\mu=0$), 
  (b) the high-doping limit ($|\mu|\gg{}\hbar{}v_F/L$).
  Results obtained from Eqs.\ (\ref{ithpdiff}) and (\ref{ithsubsh}) are
  displayed with color thick lines.
  Thin black lines visualize the tunneling limit, see Eq.\ (\ref{ijotunn})
  (dashed lines), and the ballistic limit, see Eq.\ (\ref{ijoball})
  (solid lines). 
}
\end{figure*}

\section{A~rectangular barrier and infinitely-doped leads}
\label{recbar}

For a~rectangular barrier of infinite height, corresponding to
$m\rightarrow\infty$ and $V_0\rightarrow\infty$ in Eq.\ (\ref{vxmpot}),
the transmission probability can be written in a~compact form \cite{Ryc09}
\begin{equation}
  \label{ttnrect}
  T_{k_y}(E) = \left[
    1+\left(\dfrac{k_y}{\varkappa}\right)^2\sin^2\left(\varkappa{}\,L\right)
  \right]^{-1}, 
\end{equation}
with
\begin{equation}
  \label{kncases}
  \varkappa = \begin{cases}
  \sqrt{k_F^2-k_y^2}, & \text{for }\  |k_y|\leqslant{}k_F, \\
  i\sqrt{k_y^2-k_F^2}, & \text{for }\  |k_y|>k_F, \\
  \end{cases}
\end{equation}
and the Fermi wavenumber $k_F=|E|/(\hbar{}v_F)$.

For the Dirac point ($\mu=E=0$) all transmision probabilities
$T_n(\mu)=T_{k_y}(E)$ for $k_y=k_y^{(n)}$ correspond to imaginary $\varkappa$-s
in Eq. (\ref{kncases}), indicating the transport via evenescent modes. 
Additionally, for the wide-sample limit, $W\gg{}L$,
one can approximate the summations in Eqs.\ (\ref{ijoth}) and (\ref{rnlan})
by integrations $0\leqslant{}k_y\leqslant{}k_F$, obtaining \cite{Tit06}
\begin{equation}
\label{ithpdiff}
  I(\theta)=\frac{e\Delta_0}{\hbar}\frac{2W}{\pi{}L}
  \cos(\theta/2)\mbox{artanh}[\sin(\theta/2)],  
\end{equation} 
and
\begin{equation}
\label{rnpdiff}
  R_N^{-1}=\frac{4e^2}{h}\frac{W}{\pi{}L}. 
\end{equation}
Finding the maximum of $I(\theta)$ numerically, we get
\begin{equation}
\label{icrnpdiff}
I_cR_N\frac{e}{\Delta_0}=2.0821 \ \ \ \ 
\text{and }\ \ \ 
S=0.2548. 
\end{equation}

In the high-doping limit ($|\mu|\gg{}\hbar{}v_F/L$), the quantities
considered are governed by real $\varkappa$-s in Eq. (\ref{kncases}); 
however, the argument of sine in Eq.\ (\ref{ttnrect}) can be regarded
as a~random phase ($\varphi$), the integration over which can be performed
independently form the integration over $0\leqslant{}k_y\leqslant{}k_F$.
In turn,
\begin{equation}
\label{ithsubsh}
I(\theta)\simeq{}\frac{e\Delta_0}{\hbar}
\frac{W}{\pi}\int_0^{k_F}dk_y\frac{1}{\pi}\int_0^{\pi}d\varphi 
\frac{T_{k_y,\varphi}\sin\theta}{\sqrt{1-T_{k_y,\varphi}\sin^2(\theta/2)}},
\end{equation}
where
$
  T_{k_y,\varphi} = \left[
  1+\dfrac{k_y^2}{k_F^2-k_y^2}\sin^2\varphi
  \right]^{-1}
$. 
For the normal-state resistance, the analogous integrations can be
performed analytically \cite{Ryc21b}, leading to
\begin{align}
  \left(T_{k_y}\right)_{\rm approx} &=
  \frac{1}{\pi}\int_0^{\pi}d\varphi\, T_{k_y,\varphi}
  = \sqrt{1-\left(k_y/k_F\right)^2}, \\
  R_N^{-1} &\simeq{} \frac{g_0W}{\pi}\int_0^{k_F}dk_y\big(T_{k_y}\big)_{\rm approx}
  =\frac{\pi}{4}\,G_{\rm Sharvin}, \label{rnsubsh}
\end{align}
with $g_0=4e^2/h$ (the conductance quantum for graphene) and the Sharvin
conductance 
$G_{\rm Sharvin}=g_0k_FW/\pi=g_0|\mu|W/(\pi\hbar{}v_F)$.
The prefactor of $\pi/4$ in the last expression of Eq.\ (\ref{rnsubsh}),
which was derived for a~rectangular sample, 
justifies the notion of 'sub-Sharvin conductance' and depends only weakly 
on the system geometry (see Ref.\ \cite{Ryc25}). 
The numerical optimization for Eq.\ (\ref{ithsubsh}), with respect to
$\theta$, leads to
\begin{equation}
\label{icrnsubsh}
I_cR_N\frac{e}{\Delta_0}\simeq{}2.4285 \ \ \ \ 
\text{and }\ \ \ 
S=0.4160.
\end{equation}

In Fig.\ \ref{ithrect1ab}, we display the current-phase relations given
by Eqs.\ (\ref{ithpdiff}) and (\ref{ithsubsh}) multiplied by the
relevant normal-state resistance, see Eqs.\ (\ref{rnpdiff}) and
(\ref{rnsubsh}), and by a~factor of $e/\Delta_0$ to obtain dimensionless
results (thick solid lines). 
Also in Fig.\ \ref{ithrect1ab}, the two limiting cases for the mesoscopic
Josephson equation are illustrated.

The first one is a~tunel junction, with $T_n\ll{}1$ for all $n$, for which
one can linearize Eq.\ (\ref{ijoth}) in
$T_n$,  with the result
\begin{equation}
\label{ijotunn}
  I(\theta)\simeq\frac{e\Delta_0}{2\hbar}\sum_{n=0}^{N-1}
  T_n\sin\theta = \frac{\pi\Delta_0}{2e}R_N^{-1}\sin\theta, 
\end{equation}
which is plotted with dashed lines in Fig.\ \ref{ithrect1ab}.
In such a~limit, we have
\begin{equation}
\label{icrntunn}
  I_cR_N\frac{e}{\Delta_0}=\frac{\pi}{2} \ \ \ \ 
\text{and }\ \ \ 
S=0.  
\end{equation}

The second is a~perfect ballistic system (or Sharvin contact), for which
transmission eigenvalues are equal to either $0$ or $1$, and can be order
such that
\begin{equation}
  T_n=\begin{cases} 1, & 0\leqslant{}n<N_0,\\
  0, & N_0\leqslant{}n<N, 
\end{cases}
\end{equation}
with the number of open channels $N_0\ll{}N$.
Substituting the above to Eqs.\ (\ref{ijoth}) and (\ref{rnlan}) we
immediately get
\begin{equation}
\label{ijoball}
  I(\theta)=N_0\frac{e\Delta_0}{\hbar}
  \sin(\theta/2)\,{\rm sgn}(\cos\theta/2), 
\end{equation}
where ${\rm sgn}(x)$ is the sign function, and
\begin{equation}
  R_N^{-1}=N_0\frac{4e^2}{h}. 
\end{equation}
In effect, for the ballistic limit 
\begin{equation}
\label{icrnball}
  I_cR_N\frac{e}{\Delta_0}=\pi \ \ \ \ 
\text{and }\ \ \ 
S=1.  
\end{equation}
Eq.\ (\ref{ijoball}) is visualised in Fig.\ \ref{ithrect1ab} with thin
solid lines.

As can be easily seen from Fig.\ \ref{ithrect1ab}, the graphene-specific
current-phase relation differs significantly from the familiar sine function
predicted for the tunneling limit, either at the Dirac point ($\mu=0$) or
in the high-doping limit ($|\mu|\gg{}\hbar{}v_F/L$); however, it is still far
from the relation predicted for the ballistic limit.

The influence of a finite potential step height, as well as the smooth
potential shape, is discussed next.

\begin{figure*}[t]
\includegraphics[width=0.6\linewidth]{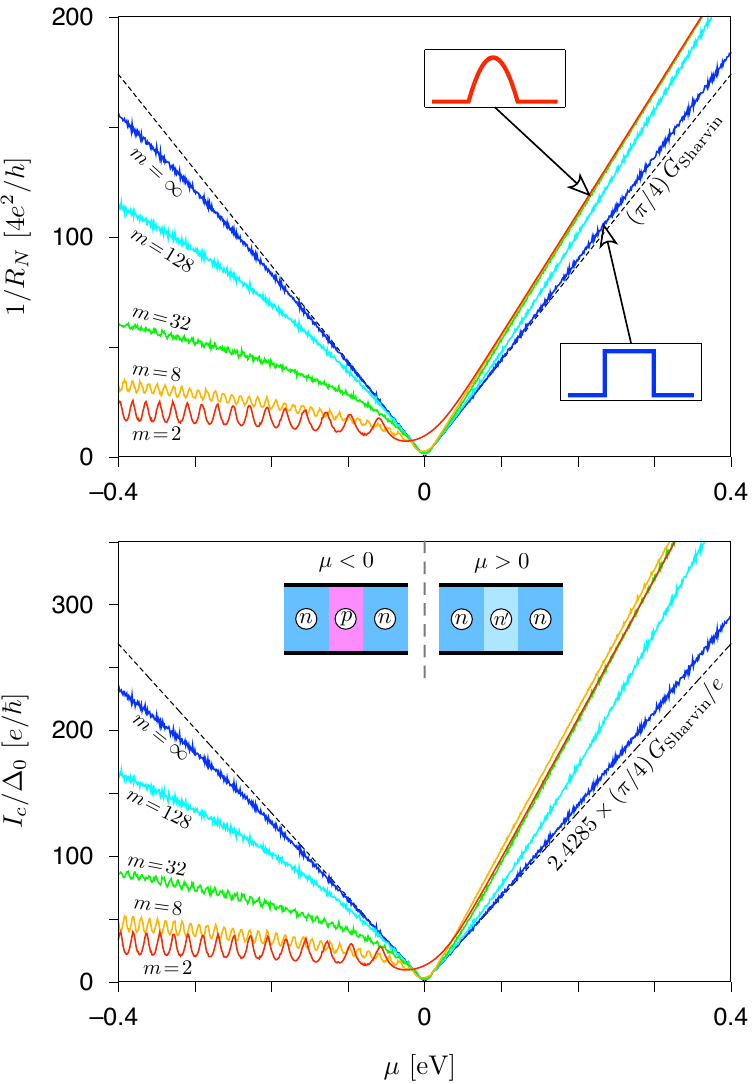}
\caption{ \label{gicV0t0ov2ab}
  Normal-state conductance $1/R_N$ (top) and critical current $I_c$ (bottom)
  for the system of Fig.\ \ref{setupjos} as functions of the chemical
  potential ($\mu=E$).  The parameters are: $W=5\,L=1000\,$nm,
  $V_0=t_0/2=1.35\,$eV. The exponent $m$ in Eq.\ (\ref{vxmpot}) is specified
  for each dataset (solid lines). Insets (top) depict the potential profiles
  for $m=2$ and $m=\infty$. Dashed line depicts the sub-Sharvin
  conductance given by Eq.\ (\ref{rnsubsh}) and the corresponding critical
  current, see Eq.\ (\ref{ithsubsh}). 
  (The values of $G_{\rm Sharvin}$
  are not shown, as they closely follow the numerical results for $m=2$
  and $\mu>0$.)
  Additional insets (bottom) visualize the tripolar (n-p-n) and the unipolar
  (n-n'-n) doping for $\mu<0$ and $\mu>0$ (respectively). 
}
\end{figure*}

\begin{figure*}[t]
\includegraphics[width=\linewidth]{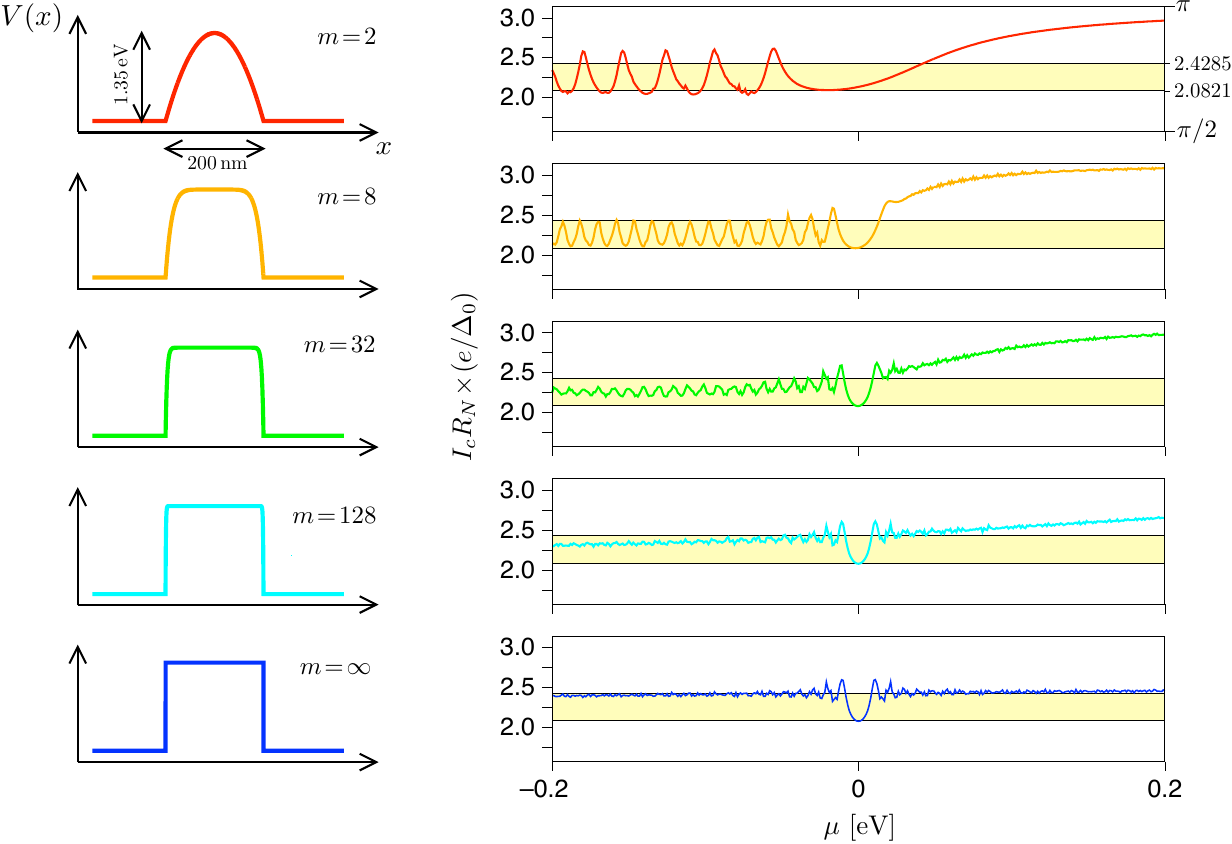}
\caption{ \label{icrnV0t0ov2}
  Product $I_cR_N$ for the data shown in Fig.\ \ref{gicV0t0ov2ab}.
  The exponent $m$ in Eq.\ (\ref{vxmpot}) is varied between the rows. 
  The potential profile for each $m$ is displayed on the left. 
  Horizonal lines bordering the yellow areas on the right mark the 
  values for rectangular barrier of an infinite height ($m\rightarrow\infty$,
  $V_0\rightarrow\infty$), corresponding to $\mu=0$ and
  $|\mu|\gg{}\hbar{}v_F/L$, see Eqs.\ (\ref{icrnpdiff}) and (\ref{icrnsubsh}). 
}
\end{figure*}

\begin{figure*}[t]
\includegraphics[width=0.6\linewidth]{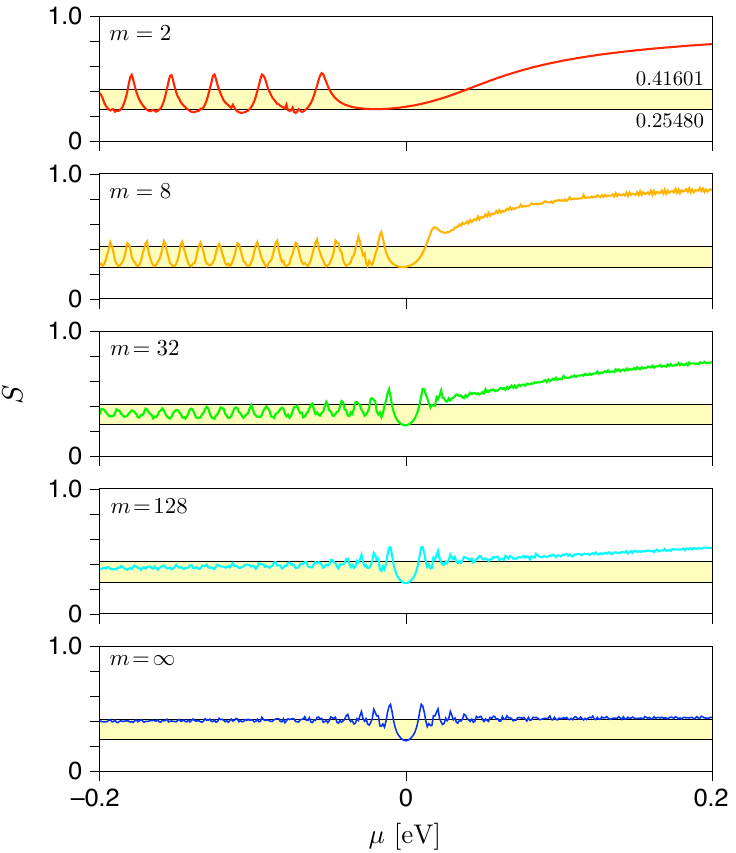}
\caption{ \label{smaxV0t0ov2}
  Skewness of the current-phase relation $S$ displayed versus the chemical
  potential for the same system parameters as in Fig.\ \ref{gicV0t0ov2ab}
  (solid lines). The exponent $m$ in Eq.\ (\ref{vxmpot}) is specified
  for each panel. Horizonal lines mark graphene-specific values of $S$, 
  see Eqs.\ (\ref{icrnpdiff}) and (\ref{icrnsubsh}). 
}
\end{figure*}

\section{Smooth potential barriers}
\label{smopot}

In this section, we present the central results of the paper, concerning the
normal-state conductance $1/R_N$, the critical current $I_c$,
the product $I_cR_N$, and skewness of the current-phase relation for
superconductor-graphene-superconductor (S-g-S) Josephson junction depicted
in Fig.\ \ref{setupjos}.
The numerical calculations are carried out according to Eqs.\
(\ref{vxmpot})--(\ref{lsysrt}), for the system with infinite-mass boundary
conditions and the width $W=5L=1000\,$nm. This sets the energy scale that 
separates the weak- and the high-doping regimes at $\hbar{}v_F/L=2.9\,$meV. 
The step height in Eq.\ (\ref{vxmpot}) is $V_0=t_0/2=1.35\,$eV, which yields
(for instance) the number of propagating modes in the leads as $N=691$ for
$\mu=-0.1\,$eV and $N=802$ for $\mu=0.1\,$eV. 
The numerical integration of Eqs.\ (\ref{phapri}) and (\ref{phbpri}) was 
performed utilizing a~standard fourth-order Runge-Kutta algorithm, with
a~spacial step of $\Delta{}x=L/800=2.5\,$pm. 
(For each value of $k_y$, the system analogous to Eq.\ (\ref{lsysrt}), but
describing scattering from the right to the left, was solved independently
to find the amplitudes $t'$ and $r'$, and to check the unitarity of the
scattering matrix $\mathcal{S}$; the above parameters result in an unitarity
error $\epsilon_n=||\mathcal{S}\mathcal{S}^\dagger-I||<10^{-8}$, with $||A||$
denoting the maximum absolute value of a~matrix element $A_{ij}$ and the
identity matrix $I$, for all cases considered.)  
Summation over the modes in Eqs.\ (\ref{ijoth}), (\ref{rnlan}) was terminated
if $T_n<10^{-6}$. 

The evolution of the conductance spectrum with exponent $m$ in Eq.\
(\ref{vxmpot}), which was discussed earlier in Ref.\ \cite{Ryc21b},
is visualized in the top panel of Fig.\ \ref{gicV0t0ov2ab}.
The bottom panel of Fig.\ \ref{gicV0t0ov2ab} presents a~similar evolution
of the critical current. 
Remarkably, the behavior of the two physical properties is very similar,  
particularly in the tripolar regime ($\mu<0$). 
For the unipolar regime ($\mu>0$), $I_c$ decays slightly faster than $1/R_N$
with the increasing $m$, exhibiting irregular behavior for smaller $m$-s. 

A~deeper insight into the system behavior is provided by the evolution
of the product $I_cR_N$, as shown in Fig.\ \ref{icrnV0t0ov2}. 
It is noticeable that the numerical results for smooth potentials and finite
$V_0$ stay close to the {\em graphene-specific} range, which is 
defined by the values obtained 
for the rectangular barrier of an infinite height for $\mu=0$ and
$|\mu|\gg{}\hbar{}v_F/L$, see Eqs.\ (\ref{icrnpdiff}) and (\ref{icrnsubsh}),
provided that the system is in the tripolar regime ($\mu<0$). 
In contrast, in the unipolar regime ($\mu>0$), $I_cR_N$ evolves
--- with the increasing $m$ --- from
the values close to the ballistic limit, see Eq.\ (\ref{icrnball}),
towards graphene-specific values. 
(For a~ballistic sample, the tunneling limit, see Eq.\ (\ref{icrntunn}),
is not approached.) 

The above observations are further supported with the
evolution of skewness ($S$) presented in Fig.\ \ref{smaxV0t0ov2}. 
Similarly as for $I_cR_N$, the numerical results for all $m$-s stay within
or very close to the graphene-specific range as long as $\mu<0$.
For $\mu>0$, the evolution from the almost balistic value towards the higher
graphene-specific value is observed as $m$ increases. 

Another striking feature of the results presented in Figs.\ \ref{icrnV0t0ov2}
and \ref{smaxV0t0ov2} is that in the vicinity of the Dirac point,
$|\mu|\lesssim{}\hbar{}v_F/L$, both $I_cR_N$ and $S$ are very close
to the values given in Eq.\ (\ref{icrnpdiff}) and are almost 
unaffected by varying $m$. 
(For $m=2$, the minima of $I_cR_N$ and $S$ are shifted towards the tripolar
regime, but the minimal values still coincide with the predictions given
in Eq.\ (\ref{icrnpdiff}.)
We emphasize that such robustness is absent for both $1/R_N$ and $I_c$, 
see Fig.\ \ref{gicV0t0ov2ab}. This suggests that intensive quantities,
such as the product $I_cR_N$ and skewness $S$, are better probes of
graphene-specific features in S-g-S Josephson junctions. 
In particular, the $S$ measurement is weakly affected by contact resistances,
and the experimental works for S-g-S junctions \cite{Eng16,Nan17} 
report values of $S\approx{}0.2\div{}0.25$ near the Dirac point as well as 
in the tripolar regime. 

\begin{figure*}[t]
\includegraphics[width=0.6\linewidth]{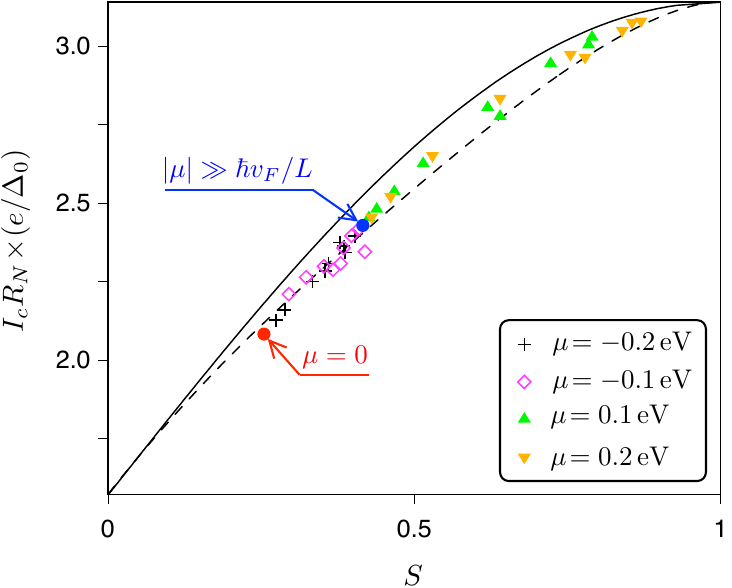}
\caption{ \label{diagIcSmpot}
  Product $I_cR_N$ displayed versus skewness of the current-phase relation
  (datapoints).
  Each dataset contains nine datapoints corresponding to $m=2,4,\dots,256$,
  and $\infty$, for a~fixed value of the chemical potential,
  $\mu=\pm{}0.1\,$eV, or $\pm{}0.2\,$eV (see the legend).
  Remaining system parameters are same as in Fig.\ \ref{gicV0t0ov2ab}. 
  Two additional symbols (full circles) mark the values given in
  Eqs.\ (\ref{icrnpdiff}) and (\ref{icrnsubsh}).
  Black solid line presents the results following from the optimization
  of Eq.\ (\ref{ijoth}) for a~single nonzero eigenvalue $0<T\leqslant{}1$; 
  black dashed line depicts the results obtaind within the toy model
  defined by Eq.\ (\ref{tttoy}). 
}
\end{figure*}

For a~somewhat more detailed view of the data we present, in Fig.\
\ref{diagIcSmpot}, the product $I_cR_N$ as a~function of skewness $S$
for four selected values of $\mu$ (since the results near the Dirac point
are almost unaffected by changing $m$, relatively high values of $\mu$ were
chosen: $\mu=\pm{}0.1\,$eV, $\pm{}0.2\,$eV)  
and nine different values of $m$ ($m=2,4,\dots,256$, and $\infty$) for each
$\mu$. 
The two pairs of $(I_cR_N,S)$ defined by Eqs.\ (\ref{icrnpdiff})
and (\ref{icrnsubsh}), bounding the graphene-specific range, are also marked 
(and indicated with arrows). 

To rationalize our numerical results for S-g-S Josephson junction,
let us first consider the case of a~single nonzero eigenvalue,
$0<T\leqslant{}1$ in Eqs.\ (\ref{ijoth}) and (\ref{rnlan}).
Straightforward optimization of $I(\theta)$ with respect to $\theta$
leads to the dependence of $I_cR_N$ versus $S$ presented with black solid
line in Fig.\ \ref{diagIcSmpot}.
It easy to see that the characteristics of such {\em single-mode Josephson
junction} are quite distant from the results for S-g-S junction. 
For this reason, 
we put forward a~toy model  that parametrizes the tunneling-to-ballistic
crossover, namely
\begin{equation}
\label{tttoy}
  T_{k_y}^{(\Theta)}=
  \frac{1}{e^{Lk_y-\Theta}+1} - \frac{1}{e^{Lk_y+\Theta}+1}
  \ \ \ \ \ \ (\Theta>0).  
\end{equation}
It is also supposed that $W\gg{}L$ and the number of modes in the leads
$N\rightarrow{}\infty$, such that the summations in Eqs.\ (\ref{ijoth}) and
(\ref{rnlan}) are replaced by integrations over $0\leqslant{}k_y<\infty$. 
For instance, $\Theta\rightarrow{}0$ reproduces the tunneling limit, with
the values of $I_cR_N$ and $S$ given by Eq.\ (\ref{icrntunn}), whereas
$\Theta\gg{}1$ corresponds to the ballistic limit. 
(To be more specific, the value of $I_cR_N$ given in Eq.\ (\ref{icrnball}) is
reproduced with an accuracy better than $1\%$ for $\Theta\geqslant{}1200$; 
the same applies for $S$, starting from $\Theta=4\cdot{}10^4$.)  

The functional dependence of $I_cR_N$ on $S$, which follows from 
Eq.\ (\ref{tttoy}), is also visualised in Fig.\ \ref{diagIcSmpot}
(dotted line). 
We find that such dependence can be approximated by
\begin{equation}
  I_cR_N\frac{e}{\Delta_0} \simeq
  \frac{\pi}{2}(S+1) + 0.59\,S^{0.89}\left(1-S\right)^{0.70}. 
\end{equation}
The corresponding curve is omitted since it matches the dotted line in Fig.\
\ref{diagIcSmpot} perfectly. 

In contrast to single-mode Josephson junction, 
the toy-model results stay very close to our S-g-S junction data obtained
for $\mu<0$ and smooth potentials. 
They also approach the values of 
$(I_cR_N,S)$ given in Eqs.\ (\ref{icrnpdiff}) and (\ref{icrnsubsh}),
with the former being well-reproduced for $\Theta\approx{}3.4$, and the
latter for $\Theta\approx{}6.8$.
The data for $\mu>0$ and smooth potentials are also not far from the
toy-model results, staying within the range of
\begin{equation}
  \theta_c < I_c{}R_N\frac{e}{\Delta_0} \lesssim 1.1\,\theta_c,  
\end{equation}
with $\theta_c=\pi(S+1)/2$.

\section{Conclusions}
\label{conclu}

We have investigated the basic characteristics of the
superconductor-graphene-superconductor (S-g-S) Josephson junction,
including the normal-state conductance, critical current, and the skewness
of the current phase relation, supposing that the potential profile along
the system is tuned from a~parabolic to a~rectangular shape. 
Compared to earlier studies of the
similar system with a~rectangular potential \cite{Tit06,Mog06}, several
new features have been identified:
(i) If the doping in the graphene sample close to the Dirac point, the product
of critical current and normal-state resistance, as well as the skewness, are
very weakly dependent on the potential shape.
This is in contrast to the critical current and the normal-state resistance
themselves, which show considerable dependence on the potential shape. 
(ii) In the tripolar regime, where the central part of the sample with hole
doping is attached to two regions with electron doping, the aforementioned
quantities take on graphene-specific values that are characteristic for an
idealized rectangular barrier of infinite height. 
(iii) In the unipolar regime (electron doping in all parts of the
sample) deviations from graphene-specific values are significant and
increase when the potential barrier is smoothed. 

Therefore, when searching for graphene-specific features in the
characteristics of S-g-S Josephson junction, the rectangular potential
profile considered in earlier studies is not essential; such features may be
observed not only near the Dirac point, but also for
relatively high dopings in the tripolar regime. 

Moreover, the approximation technique \cite{Ryc25} previously used to
calculate the conductance and shot-noise power for a~rectangular barrier
of infinite height, was adopted for the S-g-S setup to calculate the
current-phase relation, 
improving the approximations for critical current and skewness
away from the Dirac point. 

We hope that our numerical study --- if confirmed experimentally --- will extend
the list of measurable quantities with graphene-specific values, 
including the universal conductivity and
shot-noise power \cite{Kat20}, quantized visible light absorption
\cite{Nai08}, and the anomalous thermal-to-electric conductivity
ratio \cite{Cro16,Sus18,YTu23}.

\section*{Acknowledgments}
The work was partly completed during a sabbatical granted by the Jagiellonian
University in the summer semester of 2024/25. 
We gratefully acknowledge Polish supercomputing infrastructure PLGrid
(HPC Center: ACK Cyfronet AGH) for providing computer facilities and support
within computational grant No.\ PLG/2025/018544.

%%%%%%%%%%%%%%%%%%%%%%%%%%%%%%%%%%%%%%%%%%%%%%%%%%%%%%%%%%%%%%%%%%%%%%%%%%%%%%
%%%%%%%%%%%%%%%%%%%%%%%%%%%%%%%%%%%%%%%%%%%%%%%%%%%%%%%%%%%%%%%%%%%%%%%%%%%%%%

%%%%%%%%%%%%%%%%%%%%%%%%%%%%%%%%%%%%%%%%%%%%%%%%%%%%%%%%%%%%%%%%%%%%%%%%%%%%%%
%%%%%%%%%%%%%%%%%%%%%%%%%%%%%%%%%%%%%%%%%%%%%%%%%%%%%%%%%%%%%%%%%%%%%%%%%%%%%%

\end{document}